\documentclass[nonacm]{acmart}
\usepackage{balance}
\AtBeginDocument{%
  \providecommand\BibTeX{{%
    \normalfont B\kern-0.5em{\scshape i\kern-0.25em b}\kern-0.8em\TeX}}}

\setcopyright{rightsretained}
\copyrightyear{2024}
\acmYear{2024}
\acmDOI{XXXXXXX}

\acmConference[WWW '24]{WWW 2024}{May 13--17,  2024}{Singapore}

\acmISBN{978-1-4503-XXXX-X/18/06}

\usepackage{float}
\usepackage{natbib}
\usepackage{graphics}
 \graphicspath{ {images/} }

\begin{document}

% \title{Interaction Analysis of Coordinated and Organic Users: A Case Study on Tweets about the Gaza Conflict}
\title{Coordinated Activity Modulates the Behavior and Emotions of Organic Users: A Case Study on Tweets about the Gaza Conflict}


\author{Priyanka Dey, Luca Luceri, Emilio Ferrara}
\affiliation{\institution{University of Southern California} \country{Los Angeles, CA, 90007, USA}}
  % \city{Los Angeles}
  % \state{California}
  % \country{USA}
  % \postcode{90007}}
\email{deyp@usc.edu, lluceri@isi.edu, emiliofe@usc.edu}

% \author{Priyanka Dey}
% \affiliation{%
%  \institution{University of Southern California}
%  \city{Los Angeles}
%  \state{California}
%  \country{USA}
%  }
%  \email{deyp@usc.edu}

% \author{Luca Luceri}
% \affiliation{%
%   \institution{USC Information Sciences Institute}
%   \streetaddress{}
%   \city{Los Angeles}
%   \state{California}
%   \email{lluceri@isi.edu}

% \author{Emilio Ferrara}
% \affiliation{%
%   \institution{University of Southern California}
%   \streetaddress{}
%   \city{Los Angeles}
%   \state{California}
%   \email{emiliofe@usc.edu}

\renewcommand{\shortauthors}{Dey et al.}

\begin{abstract}
Social media has become a crucial conduit for the swift dissemination of information during global crises. However, this also paves the way for the manipulation of narratives by malicious actors. This research delves into the interaction dynamics between coordinated (malicious) entities and organic (regular) users on Twitter amidst the Gaza conflict. Through the analysis of approximately 3.5 million tweets from over 1.3 million users, our study uncovers that coordinated users significantly impact the information landscape, successfully disseminating their content across the network: a substantial fraction of their messages is adopted and shared by organic users. Furthermore, the study documents a progressive increase in organic users' engagement with coordinated content, which is paralleled by a discernible shift towards more emotionally polarized expressions in their subsequent communications. These results highlight the critical need for vigilance and a nuanced understanding of information manipulation on social media platforms.

\end{abstract}

\begin{teaserfigure} 
    \vspace{-.5cm}
    \centering
    \includegraphics[width=1\textwidth]
    {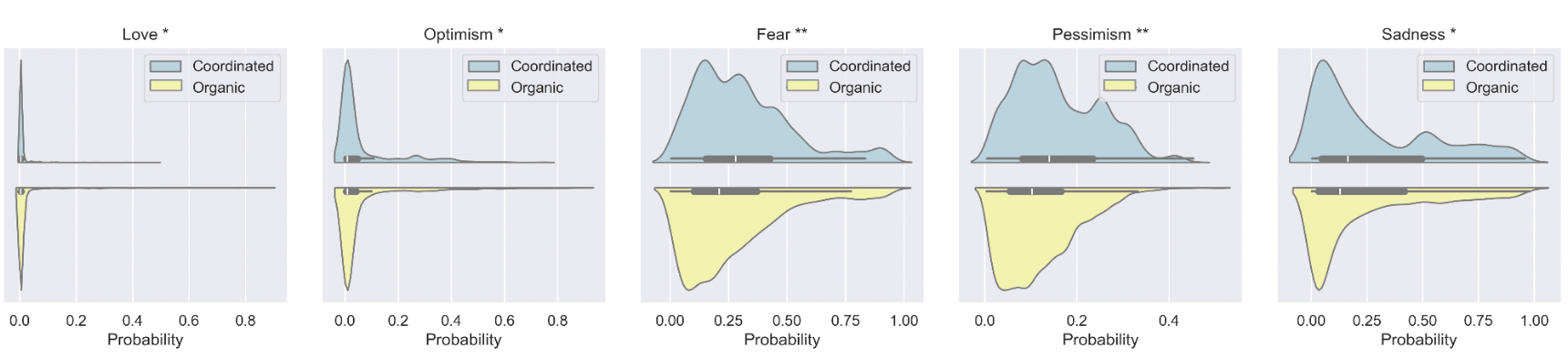}
    \vspace{-.75cm}
    \caption{Distribution of different emotions in content produced by coordinated and organic users. Each pairwise distributions' difference is significant (\textit{MW test}: * \textit{p<0.01}, ** \textit{p<0.001}). }
    \label{fig:emotions}
\end{teaserfigure}

\maketitle

\section{Introduction}

Social media has emerged as a pivotal platform for the dissemination of information on ongoing crisis events, connecting users and fostering timely information exchanges. However, the expansive reach and influence of social media also render it a potent tool for entities with malicious intentions. Through orchestrated efforts, such actors can manipulate narratives, disseminate misinformation, and advocate for particular political ideologies, thus shaping the discourse to their advantage \cite{lazer2018science, vosoughi2018spread}.

A growing body of literature has documented the exploitation of social media for such nefarious purposes, spanning from political debates to health discussions \cite{broniatowski2018weaponized, chen2021covid}. These studies elucidate how misinformation campaigns and the spread of online vitriol contribute significantly to the distortion of the digital ecosystem \cite{ref_4, ref_5,ezzeddine2023exposing}. For instance, research has illustrated the adverse effects of "fake news" on public perception and behavior \cite{ref_6, ecker2022psychological}, while others have highlighted the proliferation of misinformation and unreliable information amidst crises like the COVID-19 pandemic \cite{cinelli2020covid, chen2022charting}.

The interaction with coordinated entities and manipulated content on social media platforms has been shown to precipitate notable shifts in user behavior and psychological well-being \cite{pennycook2021psychology}. 
Exposure to misinformation has been linked with heightened anxiety and distress \cite{ref_8}, alongside a growing distrust in digital sources and a reluctance to engage in discussions on contentious issues \cite{ref_9, ref_10}.

Building on these findings, our study delves into the dynamics between coordinated and organic users on Twitter concerning the Gaza conflict, aiming to shed light on the mechanisms of information dissemination and its repercussions on user behavior.

% \vspace{-.25cm}

\subsection*{Contributions}

This research endeavors to elucidate the potential effects of coordinated content dissemination on the behavior and emotional state of organic users, guided by the following inquiries:

\begin{itemize}
\item[\textbf{RQ1}] \textit{How effective are coordinated users in disseminating their content among organic users?}
\item[\textbf{RQ2}] \textit{What are the temporal patterns of interaction between organic and coordinated users?}
\item[\textbf{RQ3}] \textit{How do organic users' behaviors and emotional expressions change subsequent to interactions with coordinated users?}
\end{itemize}

Utilizing a dataset comprising approximately 3.5 million tweets from over 1.3 million users collected from September to November 2023, we investigate the discourse surrounding the Gaza conflict. Our analysis reveals:

\begin{itemize}
\item Coordinated users effectively spread their messages through at least a tenth of the network, with over a third of their content being redistributed by organic users.
\item Engagement with coordinated content unfolds gradually, requiring sustained interaction over an extended period to observe a significant uptick in coordinated user engagement.
\item Subsequent to interactions with coordinated users, organic users exhibit a marked increase in the expression of negative emotions, notably \textit{pessimism, sadness,} and \textit{fear}. Moreover, emotions like \textit{anger} trigger a polarization among users through repeated interactions, highlighting the significant emotional and psychological effects of such engagements.
\end{itemize}

% \vspace{-.35cm}

\section{Data}
In this study of the Gaza conflict, we analyze Twitter interactions (retweets and replies) between coordinated and organic users. Our dataset covers 62 days from September 1, 2023, just before the start of the 2023 Israel-Hamas war, to November 1st, 2023. A significant tweet increase is observed after October 7th, the war's official start. We collected data using a manually curated list of  English, Arabic, and Hebrew keywords, e.g., \textit{West Bank, bombs, Gaza, Jerusalem, missiles}. Keywords like \textit{\#football, FIFA, Buckwheat, PROMO Alert, BLM, blacklivesmatter} were used to filter out irrelevant content.

We amassed 3,584,175 tweets in 57 languages, with English (93\%) most predominant, followed by Arabic (6.71\%) and Hebrew (0.04\%). As English tweets constitute over 3.3 million, we focus our analyses on them. We identify 4 types of tweets: 2,935,621 retweets, 206,663 replies, 150,997 tweets, and 40,001 quotes.

% \begin{table}[t]
  % \centering
  % \begin{tabular}{|l|c|}
  %   \hline
  %   \textbf{Statistic} & \textbf{Count} \\
  %   \hline
  %   \# of tweets & 150,997 \\
  %   \# of retweets & 2,935,621 \\
  %   \# of replies & 206,663 \\
  %   \# of quotes & 40,001 \\
  %   \hline
  % \end{tabular}
  % \caption{Aggregate statistics on English tweets (93\% of dataset)}
  % \label{tab:data_stats}
  % %\vspace.5cm}
% \end{table}

%\vspace.25cm}

\section{Methodology}

\subsection{Coordinated Activity Detection}
Various techniques have been proposed to uncover coordinated activity on social media \cite{pacheco2020unveiling, sharma2021identifying, luceri2024unmasking}. We utilize a novel technique \cite{luceri2024unmasking} to construct a user similarity network based on distinct behavioral indicators.\footnote{For more details on network creation and node pruning methodology, refer to \cite{luceri2024unmasking}} The network incorporates five behavioral traces, including sharing identical URL links, hashtags, tweet content, re-sharing the same tweets, and rapid retweeting (resharing the same tweet in less than 1 minute). Each trace contributes to a similarity network, with user similarities represented through edge weights. We then consolidate these networks into a fused graph, where links between nodes indicate connections in any individual network. To identify coordinated users, we prune nodes based on centrality, selecting those with the highest 5\% eigenvector centralities.

We detected 1,034 coordinated users. Among 100 manually analyzed users, 62 had either deleted accounts or were suspended, and 9 had protected tweets. Additionally, 68 users had small followings (< 2,000 followers), while 32 had posted over 100K times.

%\vspace.25cm}
% \vspace{-.2cm}
\subsection{Measuring Coordinated Users' Effectiveness} 

To identify the effectiveness of coordinated users, we adapt four metrics introduced by \cite{ref_12}: 

% \textit{Retweet Pervasiveness (RTP)} which measures the intrusiveness of coordinated user-generated content in organic user-generated retweets, \textit{Reply Rate (RR)} which measures the percentage of replies given by organic users to coordinated users, \textit{Organic User to Coordinated User Rate (O2CR)} which quantifies organic users' interactions with coordinated users over all the organic users' activities in the social network, and \textit{Tweet Success Rate (TSR)}: is the percentage of tweets generated by coordinated users that were retweeted more than once by an organic user.\footnote{We refer readers to \cite{ref_12} for more information on how these metrics are calculated}

\noindent \textit{Retweet Pervasiveness (RTP)} measures how often coordinated users' tweets are retweeted by organic users:
$$
RTP = \frac{\text{\# of organic retweets from coordinated users}}{\text{\# of organic user retweets}}
$$
\noindent \textit{Reply Rate (RR)} measures the percentage of replies from organic users to coordinated users' tweets:
$$
RR = \frac{\text{\# of organic user replies to coordinated users' tweets}}{\text{\# of organic user replies}}
$$
\noindent \textit{Organic to Coordinated User Rate (O2CR)} quantifies organic users' interactions with coordinated users:
$$
O2CR = \frac{\text{\# of interactions with coordinated users}}{\text{\# of organic user activities}}
$$
\noindent \textit{Tweet Success Rate (TSR)} is the percentage of coordinated users' tweets retweeted more than once by organic users:
$$
TSR = \frac{\text{\# of tweets retweeted > 1 by organic users}}{\text{\# of coordinated tweets}}
$$

% \noindent \textit{Retweet Pervasiveness (RTP)} measures intrusiveness of content by coordinated users in retweets by organic users. 
% $$
% RTP = \frac{\text{\# of organic user retweets from coordinated user tweets}}{\text{\# of organic user retweets}}
% $$
% \noindent \textit{Reply Rate (RR)} measures the percentage of replies given by organic users to coordinated users.
% $$
% RR = \frac{\text{\# of organic users' replies to coordinated users' tweets}}{\text{\# of organic users' replies}}
% $$
% \noindent \textit{Organic User to Coordinate User Rate (O2CR)} quantifies organic users' interactions with coordinated users over all the organic users' activities in the social network.
% $$
% OC2R = \frac{\text{\# of organic users' interactions with coordinated users}}{\text{\# of organic users' activity}}
% $$

% \noindent \textit{Tweet Success Rate (TSR)}: percentage of tweets by coordinated users that were retweeted more than once by an organic user: 
% $$
% TSR = \frac{\text{\# of tweets retweeted > 1 by an organic user}}{\text{\# of coordinated tweets}}
% $$

%\vspace.5cm}
% \subsection{Interactions Dynamics between Organic and Coordinated Users Over Time}
\subsection{Characterizing User Interaction Dynamics}

To measure interaction changes, we analyze how the content distribution of both organic and coordinated users shifts after 1, 2, and 3 interactions. 
% To do this, we compute proportions of intra-group (organic $\rightarrow$ organic, coordinated $\rightarrow$ coordinated) and inter-group (organic $\rightarrow$ coordinated, coordinated $\rightarrow$ organic) interactions for each user. 
To do this, we calculate the mean inter-group (organic $\rightarrow$ coordinated, coordinated $\rightarrow$ organic) and intra-group (organic $\rightarrow$ organic, coordinated $\rightarrow$ coordinated) interaction proportions for all activity between the $k$-th interaction and $k+1$-th interaction ($k \in \{1,2,3\}$). For the third interaction, we consider \textit{all} activity until November 1, 2023. 

To further observe changes in organic users' behaviors, we examine their inter-group interaction proportions after 1, 2, and 3 interactions ($k \in \{1,2,3\}$) and across different time windows ($t \in \{\textit{1 day}, \textit{3 days}, and \textit{1 week}\}$). 
% We calculate the inter-group interaction proportions for each interaction step ($k$) and time window ($t$). 
For instance, for $k=1$ and $t = \textit{1 hour}$, if a user has their 1st interaasction at 11AM and 2nd interaction at 11:58AM, all interactions between 11:00 AM and 11:58 AM are considered. We then compute the proportion of interactions with coordinated users and denote this as $\textit{O2C prop}$. Similarly, for all other organic users, we compute $O2C prop$ for $k=1, t=\textit{\text{1 day}}$. We then compute the mean $O2C prop$ over all users. We repeat this process for the remaining interaction steps and time windows.

% \vspace{-.25cm}
\subsection{Organic Users' Behavior After Interactions}
To analyze changes in user behavior, we examine original content (tweets) posted by organic users after interacting with coordinated users. We investigate variations in expressed emotions across different time windows ($t$) and interaction steps ($k$). We leverage \cite{ref_emotion_model}, which analyzes text and assigns probabilities (ranging from 0 to 1) for 11 emotions: \textit{anticipation, joy, love, optimism, surprise, trust, anger, disgust/contempt, fear, pessimism}, and \textit{sadness}.

To establish a baseline for examining shifts in the organic users' content, we first explore differences between coordinated and organic users. We gauge the significance of these deltas by using Mann-Whitney (MW) tests ($\alpha = 0.01, 0.001$). To examine content shifts across interaction steps, we identify tweets before and after the $k$-th interaction within a specified time window, $t$. For each interaction step and time window, we compute emotion deltas by averaging the difference between post-probabilities (emotion probabilities for tweets written \textit{after} the interaction step) and pre-probabilities (emotion probabilities for tweets written \textit{before} the interaction step) across all tweets and users. %MW tests with an $\alpha$-value of 0.01 assess significant differences.

% \vspace{-.25cm}

\section{Results} 

Next, we examine how successful coordinated users are in spreading their content through the network (RQ1), how the interactions between coordinated and organic users change over interaction steps (RQ2), and finally whether there are changes in users' content after interactions with coordinated content (RQ3). 

% %\vspace1.5cm}

% \vspace{-.25cm}
\subsection{Effectiveness of Coordinated Activity} 

We estimate the effectiveness of the 1,034 coordinated users (identified through our fused network) in receiving engagement and endorsement from organic users in the discussion related to the Gaza conflict. We identified 1,766 organic users that interact (retweet or reply) with coordinated users, and 1,326,695 users that do not have any interactions with coordinated users. Using the four metrics outlined in \S 3.2, we observed the following values: \textit{RTP} (10.51\%), \textit{RR} (10.62\%), \textit{O2CR} (9.87\%), and \textit{TSR} (36.58\%). 

We observe that organic users tend to reply to coordinated users slightly more often than they retweet them. While less than 10\% of the organic users' interactions involve coordinated users (\textit{O2CR}), over a third of coordinated user-generated  content has some engagement (\textit{TSR}), indicating substantial dissemination of coordinated content throughout the network. In comparison to \cite{ref_10}, we note lower \textit{RTP} and \textit{RR}, yet comparable \textit{O2CR} and \textit{TSR}. This illustrates that although organic users mainly engage with each other, a significant portion of coordinated content permeates various segments of the network, including vulnerable organic audiences.

\begin{figure}[t]
    %\vspace.25cm}
    \centering
    \includegraphics[width=\columnwidth]{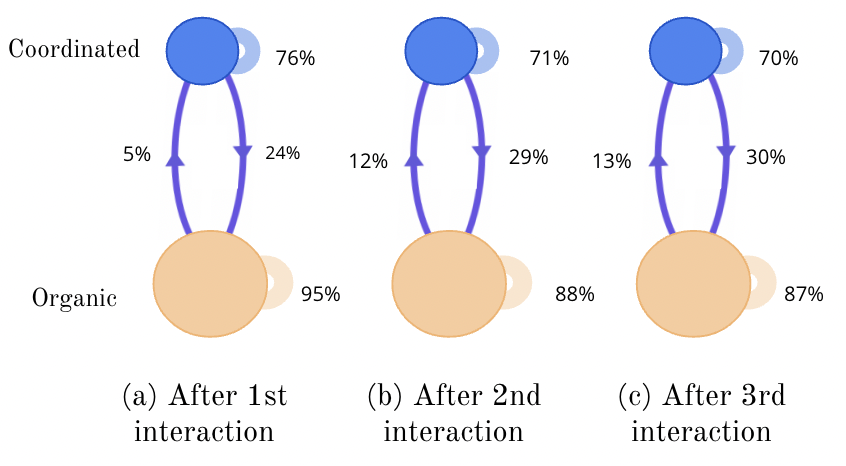}
    \vspace{-.75cm}
    \caption{Inter and intra-group engagement proportions between organic and coordinated users after $k$ interactions}
    \vspace{-.5cm}
    \label{fig:rq2_markov}
\end{figure}

%\vspace.25cm}
\subsection{Interactions Increase Over Time} 

In this section, we examine the evolving interactions between organic and coordinated users over time. Figure \ref{fig:rq2_markov} illustrates the shifting content distributions of both groups over interaction steps. 
As the number of interactions rises, we observe descending trends in intra-group interaction proportions. However, we also see a rise in inter-group interaction proportions, indicating the ability of coordinated users to initiate discussions with organic users through repeated interactions.

As we are interested in understanding the organic users' behavior, we further explore their changes in interaction proportions over three time windows (\textit{1 day}, \textit{3 days}, \textit{1 week}) by averaging the metric \textit{O2C prop} over all users as discussed in \S 3.3. Results from this analysis suggest
% are presented in the line plot in Figure \ref{fig:rq2_lineplot}. We observe 
that within a week after 3 interactions, over 40\% of organic user interactions are with coordinated users. These results further signal that organic users may be attracted to the content posted by coordinated users, thus leading to a larger number of interactions with them.

% \begin{figure}[h]
%     \centering
%     \includegraphics[width=0.4\textwidth]{images/rq2_lineplot.png}
%     %%\vspace10pt}
%     \caption{Proportion of Organic $\rightarrow$ Coordinated Interactions Over Varying Time Intervals and Interaction Steps}
%     \label{fig:rq2_lineplot}
% \end{figure}

% \subsection{RQ3: Change in Organic Users' Content After Interactions Over Time}

%\vspace.25cm}

\subsection{Emotion Modulation by Coordinated Users}

Our third RQ investigates whether repeated interactions alter organic users' content. We compare emotions expressed in coordinated and organic content in Figure \ref{fig:emotions}. We present the distributions of five emotions (2 positive: \textit{Love, Optimism} and 3 negative: \textit{Fear, Pessimism, Sadness}), which significantly differ (MW test, $\alpha$ = 0.01). 
Our findings indicate that coordinated users are less likely to incorporate emotions such as \textit{love} and \textit{optimism} in their content, as evidenced by the narrower probability ranges, namely (0 to 0.5 and 0 to 0.8), in contrast to organic users.
Comparable patterns emerge for the other positive emotions (\textit{Anticipation, Joy, Surprise, Trust}). However, coordinated content tends to exhibit higher probabilities of emotions such as \textit{Anger} and \textit{Disgust/Contempt},  suggesting more negative emotion usage overall.
% %\vspace.25cm}

\begin{figure}[t]
    \centering
    \includegraphics[width=.96\columnwidth]{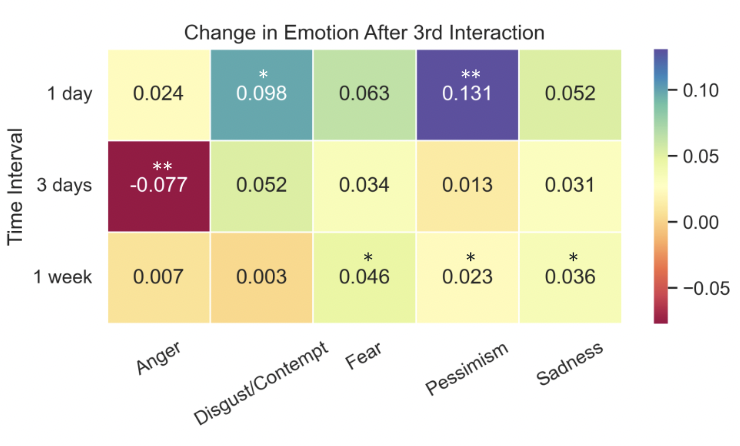}
    \vspace{-.6cm}
    \caption{Changes in negative emotions after 3 interactions. Each change is  significant (\textit{MW test}: * \textit{p<0.01}, ** \textit{p<0.001}).}
    \label{fig:rq3_heatmap}
    \vspace{-.5cm}
\end{figure}

Given that the emotions used by the two groups are significantly different, we study the emotions expressed in organic user content after interacting with coordinated users. 
% justification for including only 3rd interaction results
As results from RQ2 suggest that changes in user behavior are most prevalent after the 3rd interaction, we illustrate variations in negative emotions following the third interaction in Figure \ref{fig:rq3_heatmap}.

% moving this to methodology
% ; we measure differences in emotions after different interaction steps ($i \in \{1, 2, 3]\}$) and time intervals ($t \in \{\textit{1 day}, \textit{3 days}, \textit{1 week}\}$). We compute the deltas in each emotion by averaging the difference between the post-probability and the pre-probability across all tweets for each user. 

% In Figure \ref{fig:rq3_heatmap}, we illustrate variations in negative emotions following the third interaction. As our analysis suggests that the most significant changes tend to manifest after three steps, we present results from the 3rd interaction.
% \footnote{We denote significant differences ($\alpha= 0.01$) with asterisks} 
We observe a steady increase in the use of a majority of the negative emotions (as signaled by positive delta values). 
%The only emotion that experiences a decrease is \textit{Anger} within the 3 days window. %
Although deltas show that after interacting with coordinated users, organic users use even less positive emotions, many of these differences are not significant. For interactions 1 and 2, 
% Finally, due to limited space, we exclude plots from 1st interactions and 2nd interactions. 
we observe some similar patterns: negative emotions tend to be more significant in content written after interactions and as the time window increases, we observe more significant differences. We also notice some fleeting emotions such as \textit{trust}, which reduces after 2 interactions during the \textit{1 day} and \textit{3 days} time intervals, but after \textit{1 week}, there are no significant differences. 

\begin{figure}[t]
    \centering
    \includegraphics[width=\columnwidth]{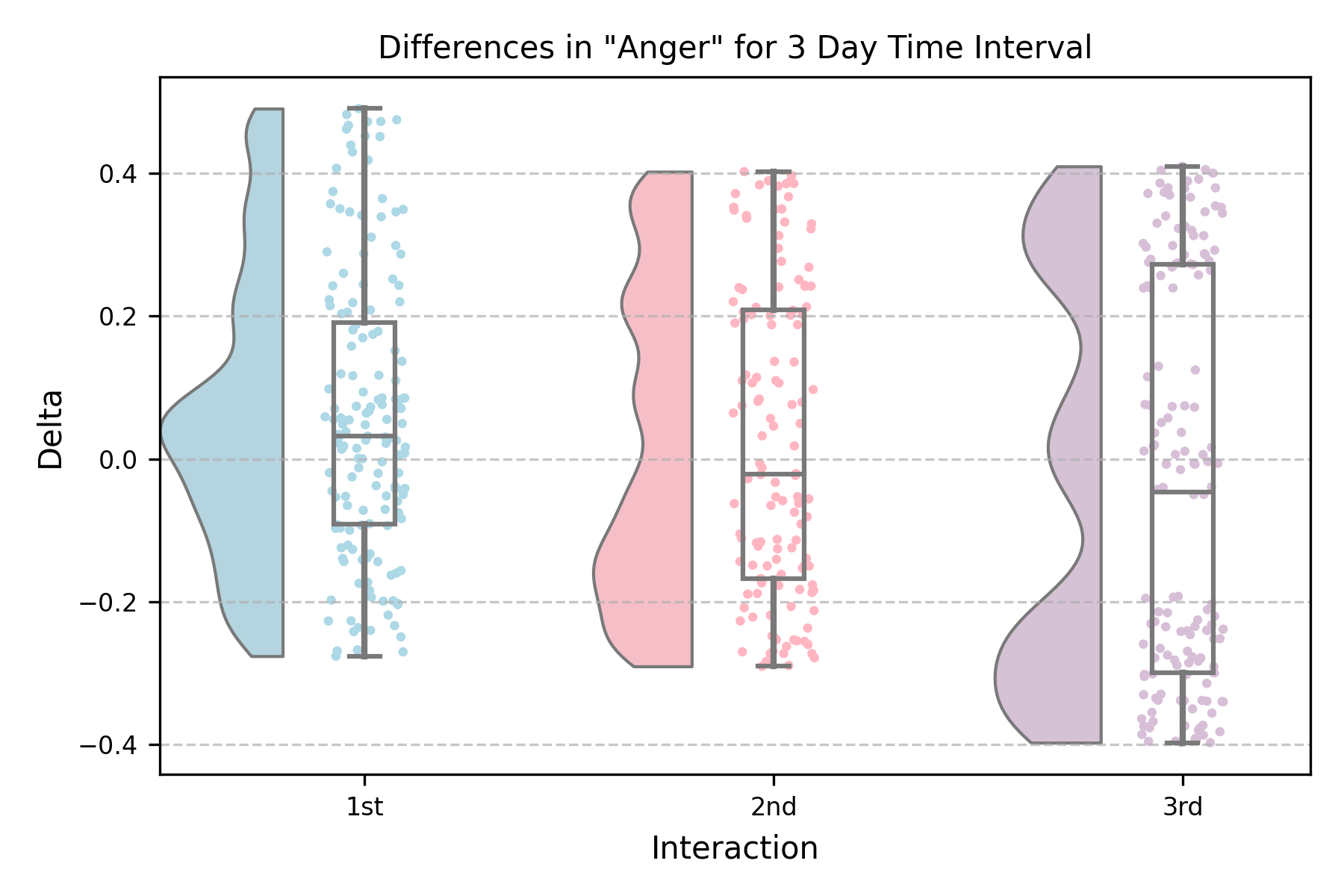}
    \vspace{-1cm}
    \caption{Distribution of \textit{Anger} deltas for \textit{3 days} interval over repeated interactions between organic and coordinated users. Noteworthy, the \textit{Anger} deltas become increasingly polarized, suggesting that repeated interactions with coordinated activity leads to increasingly extreme changes in \textit{Anger} emotion.}
    \label{fig:anger_distribution}
    \vspace{-.25cm}
\end{figure}

An intriguing observation we noted is the decrease in the use of \textit{anger} within the \textit{3 days} time window following the third interaction, followed by an increase in usage after \textit{1 week}. Similar patterns emerge after 1 and 2 interactions. To delve deeper into this phenomenon, we analyze the distribution of delta values in \textit{anger} after each interaction within the \textit{3 days} interval (refer to Figure \ref{fig:anger_distribution}).

Upon closer examination of the plots, we observe that repeated interactions pushed users towards the extreme. Users diverge into three distinct groups over repeated interactions within a brief three-day span, with two displaying notable radical tendencies. One group shows increased anger (higher delta values) after interactions, while the other tends to become more subdued (lower delta values). 

We manually reviewed some user content to illustrate this trend. For instance, one user posted: \textit{"... \#FoxNews Hmm don't remember you uttering a word Bernie when Hamas slaughtered civilians and beheaded infants. Selective faux outrage u commie!"} after the 1st interaction, but later shared \textit{"Gaza crisis: Angelina Jolie's heartfelt post for peace..."} after the 3rd interaction, indicating a decrease in anger. Another user posted \textit{"Defund the universities! We can't ignore Jew-hating academia..."} after the 1st interaction and \textit{"... Seriously.... what the **** do they know about Israel/Palestine - Republican sympathising Teachers and parents pushing their agenda!"} after the 3rd interaction, demonstrating a significant increase in anger.
%\vspace{-.5cm}

Our analysis suggests that interactions with coordinated users often shape how organic users create new posts and express emotions within their messages. A notable portion of their content tends to exhibit heightened negative emotions. Interestingly, concerning specific emotions like \textit{anger}, users form distinct groups over a period of time, with some showing increased anger while others seem to become more desensitized.

%\vspace.25cm}

\section{Conclusions}

In summary, our analysis highlights the potential effects of coordinated users on social media dynamics. They effectively disseminate messages, impacting a substantial portion of the network, with a notable fraction of their content redistributed by organic users. Although engagement with coordinated content unfolds gradually and requires sustained interaction over time, the interactions can have behavioral and psychological effects on organic users. New content posted by these users express heightened negative emotions, including \textit{pessimism, sadness,} and \textit{fear}, while emotions like \textit{anger} drive polarization among users. These findings underscore the need for further research to explore shifts in users' behaviors post-interactions, with investigating topics and linguistic style changes emerging as compelling avenues for future study.

% \begin{acks}
\smallskip \small \textbf{Acknoledgements}. This work was supported in part by DARPA (contract no. HR001121C0169).
% \end{acks}

%\vspace.25cm}

% \bibliographystyle{ACM-Reference-Format}
% \balance
\bibliographystyle{abbrvnat}
\bibliography{sample-base}

\end{document}